# High-resolution 4D acquisition of freely swimming human sperm cells without staining

Gili Dardikman-Yoffe, Simcha K. Mirsky, Itay Barnea, and Natan T. Shaked*

Department of Biomedical Engineering, Faculty of Engineering, Tel Aviv University, Tel Aviv 69978, Israel

**Abstract**
We present a new acquisition method that enables high-resolution, fine-detail full reconstruction of the three-dimensional movement and structure of individual human sperm cells swimming freely. We achieve both retrieval of the three-dimensional refractive-index profile of the sperm head, revealing its fine internal organelles and time-varying orientation, and the detailed four-dimensional localization of the thin, highly-dynamic flagellum of the sperm cell. Live human sperm cells were acquired during free swim using a high-speed off-axis holographic system that does not require any moving elements or cell staining. The reconstruction is based solely on the natural movement of the sperm cell and a novel set of algorithms, enabling the detailed four-dimensional recovery. Using this refractive-index imaging approach, we believe we have detected an area in the cell that is attributed to the centriole. This method has great potential for both biological assays and clinical use of intact sperm cells.

## Introduction

In intracytoplasmic sperm injection (ICSI), a single sperm is selected for direct injection into the ovum. This process poses an innate difficulty, as the natural sperm selection induced by the race to the ovum, perfected by millions of years of evolution, is replaced by a selection performed by a clinician (*1*). Today, clinicians base the process of sperm selection on morphology and dynamics assessment, which is carried out using two-dimensional imaging methods, such as bright-field microscopy (BFM), differential interference contrast microscopy (DIC), Hoffman's microscopy, and Zernike's phase contrast microscopy (*2*,*3*). These imaging methods lack quantitative contrast, due to the prohibition of using exogenous contrast agents in human ICSI, and have distinctive imaging artifacts. Moreover, both the structure and the natural movement of the human sperm cell are three-dimensional (3-D); thus, an entire aspect of the dynamics and morphology of sperm cells is currently disregarded due to technological barriers. Holographic imaging allows for a fully quantitative measurement of the cell optical thickness [i.e., the product of the refractive index (RI) and the physical thickness] (*4*). This enables the stain-free examination of the sperm, based not only on the two-dimensional (2-D) morphology of the cell head, but also on optical-thickness-based parameters, such as dry mass, and phase-volume (*5*,*6*).

In addition to head morphology, sperm flagellum defects have been postulated to be prognostic for fertilization failure (*7*). However, the complex shape and motion of the flagellum is typically disregarded in sperm selection protocols due to its low visibility and 3-D nature (*1*). Most commonly, four-dimensional (4-D) tracking methods for sperm cells have been suggested based on pinpointing the 3-D location of a single point on the sperm, usually the sperm-head centroid, over time (*4, 8-11*). Corkidi et al. (*8*) suggested such a method, with a system based on a piezo-electric device displacing a large focal distance objective mounted on a microscope to acquire dozens of image stacks per second, where each stack of images spans across a depth of 100 μm. However, in this previous work, sperm trajectories were obtained manually as the main focus was the high-speed acquisition of 3-D data, thus making it inappropriate for clinical use. Su et al. (*9*) suggested an automated, high-throughput tracking method based on holographic lens-free shadows of sperm cells that are simultaneously acquired at two different wavelengths, emanating from two sources placed at 45° with respect to each other, where the 3-D location of each sperm is determined by the centroids of its head images reconstructed in the vertical and oblique channels. Digital holographic microscopy images have also been previously used to track sperm cell trajectories, by using wavefront propagation to obtain multiple focus planes, and either Tamura's coefficient (*10*), a criterion based on the invariance of both energy and amplitude (*11*), or a criterion based on the steepest intensity gradient along the *z* axis (*12*), to estimate the in-focus distance.

In the last couple of years, several studies aimed at tracking the location of the entire flagellum in 3-D space and time (*13-15*). Silva-Villalobos et al. (*13*) used an oscillating objective mounted in a bright-field optical microscope, covering a 16 µm depth at a rate of 5000 images per second, where the best flagellum focused sub-regions were associated to their respective *z* positions. Bukatin et al. (*14*) developed an algorithm that reconstructs the vertical beat component from 2-D BFM images, by first identifying the projected 2-D shape of the flagellum based on the pixel intensity levels, and then estimating the *z* coordinate by analyzing the intensity profile along cross sections normal to the flagellum in the image plane, exploiting the fact that the width of the halo is correlated to the *z* displacement from the focus plane. Finally, Daloglu et al. (*15*) recently extended their former work using simultaneous illumination from two sources emerging from two oblique angles (*9*), reporting a high-throughput and label-free holographic method that can simultaneously reconstruct both the full beating pattern of the flagellum and the translation and spin of the head. These methods evaluate the trajectories of the sperm in space, but do not address the problem of reconstructing the 3-D morphology of the head, which contains clinically significant organelles, preventing full 4-D morphological evaluation.

Label-based confocal fluorescence microscopy can obtain high spatial-resolution 3-D cell imaging (*16*). However, it cannot be used in human sperm cell selection during ICSI as staining is not allowed. Furthermore, even when using label-based confocal fluorescence microscopy for live cells in animal fertilizations or in biological studies, it is very challenging to obtain the entire sperm head and flagellum 3-D reconstruction during rapid free swim, due to the need to scan over time and the low amounts of fluorescence emission in each temporal frame. The gross surface morphology of the bovine sperm head was previously reconstructed without staining by using a Gaussian beam inducing self-rotation of the cell and the shape from silhouette algorithm (*17*). This method, however, could not capture the internal organelles of the sperm head or its 4-D flagellum beating. In tomographic phase microscopy, interferometric projections of the specimen are processed and positioned in the 3-D Fourier space, so that the RI map can be reconstructed. Recently, a commercial tomographic microscope was used to recover the 3-D RI distributions of animal (non-human) sperm cells (*18*). Yet, this method scans only 30 illumination angles, and at a limited angular range; thus it is unable to fully image freely swimming sperm cells. Thus, the problem of reconstructing the 3-D dynamic morphologies of uninterrupted swimming sperm cell heads has not yet been solved. In this paper we present the first 4-D acquisition method for the entire sperm (head, organelles and flagellum) during free swim and without staining

## Results

Our method is based on rapid quantitative phase microscopy and a novel set of reconstruction algorithms providing the highly-detailed 4-D distribution of the sperm cell during swim. Holographic videos of sperm cells were acquired by off-axis holographic phase microscopy, which allows high-resolution complex wavefront acquisition from a single camera shot of dynamic sperm cells. The acquisition system contained an inverted microscope, equipped with a 100× oil-immersion objective, 328× total magnification through the system, an external nearly-common-path interferometric module, and a fast digital camera, recording at 2000 frames per second. Sperm cells swam freely in a perfusion chamber (see Methods Section 1). We reconstruct the recorded sperm head and the flagellum individually, treating them as separate entities due to their physical dissimilarity, as described below.

### Reconstructing the 3-D morphology of the sperm head

To recover the 3-D RI profile of the sperm head, we used a tomographic phase microscopy approach that takes into consideration diffraction effects. To position the processed projections in the 3-D Fourier space, it is essential to know the angle at which the projection of the object was captured. Existing tomographic approaches are based on either illumination rotation (*18*) or controlled sample rotation (*19*), allowing a-priori knowledge on the angles of the acquired interferometric projections. Alternatively, random rotation of cells while flowing along a microfluidic channel or perfusion chamber can be used for tomography for a degenerate case, assuming rotation around a single axis (*20*, *21*). In the case of a human sperm cell swimming freely, though, the swimming pattern is too complex to apply any of the existing tomographic algorithms, as the algorithm must take into account the rotation around all three major axes: pitch, roll and yaw. Modeling the sperm cell head as an ellipsoid, we were able to build a highly-accurate algorithm for retrieving all three angles from each frame, based on the minor and major radii extracted from the segmented phase maps of the head, requiring only a single acquisition per frame. This is done by extracting the three model constants, *A*, *B* and *C*, illustrated in Fig. 1A, from the major and minor radii vectors, retrieved from hundreds of consecutive frames, while relying on the periodicity of sperm rolling. By virtue of the periodic swimming pattern, we obtain full (spiral) angular coverage for tomography, rather than the standard ±70° obtained in the illumination rotation method. Following this, each specific frame orientation is determined via interpolation, which is possible due to the basic asymmetry of the sperm cell, inducing $A \neq B \neq C$ (see further details in the Methods Section). Using this approach, we are able to recover the orientation that yields the same projected ellipse as the rotated ellipsoid model. Utilizing continuity constraints, we are able to reliably limit the cases in which there is more than one orientation that causes the same projected ellipse, and gain definitive recovery.

We then position the Rytov field of each projection at the calculated orientation in the 3-D Fourier space, allowing reconstruction of the 3-D RI distribution through optical diffraction tomography (ODT) theory (*22*). A representative result, obtained by applying our approach on 1000 recorded frames of a sperm swimming freely in a perfusion chamber, is given in Fig. 2. This result agrees with previous studies regarding the internal structure of the sperm cell head [e.g., (*23*)], showing clearly the three main components of the sperm head: cell membrane (light purple), acrosomal vesicle (yellow), and nucleus (red). Significantly, although we could not verify it experimentally, the sperm head reconstruction in Fig. 2 also reveals a higher-RI region in the proximal part of the head, closest to the midpiece, which may be attributed to the sperm centriole area [colored in dark purple in Fig. 2B]; this organelle, which is crucial for the development of the embryo after fertilization, has never been imaged before in live cells.

*Validation of the ellipsoid model for sperm-head orientation recovery*

To confirm the validity of the ellipsoid model for sperm head orientation recovery, we took as inputs both the external 3-D shape of an actual sperm cell (Fig. S1A), and its ellipsoid model (Fig. S1B), and calculated the projections of each for the roll and pitch angles retrieved for each frame (Fig. S2). Following this, we calculated the major and minor axis radii of each projection, as well as its orientation (i.e. native yaw angle). The results of this comparison are shown in Fig. S3. Fig. S3A shows the results obtained for the major radius from the projection of the actual sperm cell (cyan line) and the ellipsoid model (green circles) vs. the original measurement from the holograms (blue asterisk). The mean absolute percentage (MAP) error between the result obtained from the actual sperm cell and the ellipsoid model is 2.73%. Fig. S3B shows the results obtained for the minor radius from the projection of the actual sperm cell (cyan line) and the ellipsoid model (green circles) vs. the original measurement from the holograms (blue asterisk). The MAP error between the result obtained from the sperm cell and ellipsoid model is 1.98% after removing the segmentation bias, indicating the nearly identical shape of all vectors. Finally, Fig. S3C shows the results obtained for the native yaw angle from the projection of the actual sperm cell (blue line) and ellipsoid model (red dashed line). These results demonstrate that this set of parameters is mapped in a nearly-identical manner from both shapes. Thus, we are able to recover the 3-D orientation of an actual sperm cell using this data-derived ellipsoid model.

*Validation of the quality of the tomographic reconstruction based on the orientations available during free swim*

In order to verify that the actual swimming patterns of sperm cells allow access to a large enough part of the 3-D spatial Fourier space of the sample to yield a reliable reconstruction, we created a 3-D RI distribution simulating a sperm cell (Fig. S4A), and used it to produce phase maps as viewed from the experimentally recovered orientations (Fig. S2), which were similar for all sperm cells tested. For reference, we also tested the reconstruction yielded by a single-axis 360° sample rotation at 5° increments. These 3-D reconstruction results are given in Figs. S4(B-C). The MAP error obtained is 0.09% for the single-axis 360° reconstruction, and only 0.11% for the recovered orientations, demonstrating that the angular coverage obtained for the experimental data is enough for high-quality tomographic reconstruction.

**Reconstructing the sperm head flagellum**

To recover the 3-D location of the flagellum, which spans over various axial locations in each frame, we used each instantaneous off-axis digital hologram to reconstruct the optical wavefront at different distances from the plane of acquisition using the Rayleigh–Sommerfeld back-propagator. Then, we pinpointed the *z* location of each pixel associated with the sperm flagellum by finding the narrowest, most steep area of the flagellum in the phase-image-based z-stack, for a section perpendicular to the flagellum orientation at that pixel. Applying this well-known digital-propagation holographic principle to locate the various focus points of the flagellum is not trivial, due to the fact that the flagellum is as thin as the resolution limit, and is masked by coherent noise. We therefore developed a 4-D segmentation method, based on tracking the flagellum from its proximal to distal end for each off-axis hologram, by estimating the current segment direction and following it iteratively. In this 4-D segmentation process, we took advantage of the fact that under high framerate the flagellum location may only vary by a small increment between consecutive frames. Thus, for each frame, we used the segmentation result from the previous frame to create a weighted probability map, where pixels associated with the previous frame and their closest neighbors are of the highest likelihood to be associated with the current frame as well, and L2-norm farther pixels quickly drop to zero probability, preventing deviation from the flagellum. Further details are given in the Methods Section. A representative result obtained by applying this approach is shown in Fig. 3 and Movies S1-2. Figure 3A and Movie S1 show the focusing algorithm in action, where the vertical segment moves along the flagellum from the neck to the distal end, finding the ultimate focus plane for each location it is in. Figure 3B and Movie S2 show a 2-D representation of the resultant 3-D segmentation map of the flagellum, where the colormap indicates the recovered depth (relative to the original focus plane). Finally, Fig. 3C shows the segmentation result in three dimensions.

In the analysis described above, we treated the entire flagellum as a single, pixel-thin unit. However, the portion of the flagellum closest to the sperm head, the midpiece, is a thicker region, composed of a central filamentous core encircled by many mitochondria. The cylindrical axisymmetrical morphology of the midpiece, and the fact that it cannot be regarded as a single rigid body together with the sperm head, do not allow for the same orientation recovery principles applied for the sperm head to be applied here. We therefore divided all flagellum pixels into two groups: midpiece region and non-midpiece region, and calculated a global average (over all frames) of the maximal phase value and width of the section for each group. These average values were then used to calculate the integral (thickness-averaged) RI for each region (*24*), under the assumption that each region is a cylinder, such that the thickness of the central pixel is equal to the width of the section. Further details are provided in Methods Section 3.5.

Finally, the full dynamic reconstruction of the entire sperm cell can be attained by combining the 4-D segmentation results for the flagellum with the 3-D RI reconstruction of the head and 3-D orientation of the head retrieved for each frame. Figure 4 presents the full 4-D reconstruction results, obtained from 1000 frames recorded during free swim. Voxels, attributed to the midpiece or the rest of the flagellum, are represented by discs of constant diameter, which is equal to the calculated average width of the respective area. Figure 4A shows 4 frames (each in a different color) from the 3-D motion in Movie S3, and Fig. 4B shows 15 frames from the same 3-D motion movie, where frames of earlier times are more transparent. Figure 4C shows a single frame from the 3-D motion revealing the internal structure of the sperm cell, for the same cell shown in Figs. 2 and 3. The integral RI was found to be 1.383 in the midpiece, and 1.365 in the rest of the flagellum. Figure 4D shows the 3-D trajectory of the sperm cell head centroid, changing from blue to red as time progresses. The path shown in Fig. 4D falls under the category of the "typical" trajectory (*9*) – the most prevalent swimming pattern observed among human sperm cells (>90% of cells) – where the sperm head moves forward swiftly along a slightly curved axis with a small arbitrary lateral displacement (approximately 4 μm side-to-side) in either direction orthogonal to the flow.

We have imaged several more sperm cells swimming freely, resulting in similar 3-D RI distributions. Results for two additional sperm cell heads are given in Fig. S5.

**Discussion**

We presented a highly-detailed, stain-free, 4-D imaging method for capturing the entire sperm cell during free swim, including both the head with its internal organelle morphologies and the flagellum, based on a single-channel holographic video. Other than the clear benefit for clinicians, allowing 4-D visualization for a more informed selection of sperm cells, this reconstruction enables calculation of the volume and dry mass for each of the organelles, thus adding previously inaccessible quantitative parameters for clinical use or biomedical assays.

The suggested method assumes progressive sperm cells with typical repetitive rotation of the sperm head, including continuous unidirectional rolling, where a single cell crosses the field of view each time. Sufficient acquisition framerate is crucial both for applying the temporal proximity principle in the 4-D segmentation of the flagellum, and for properly recovering the 3-D orientation of the sperm head. For 4-D segmentation of the flagellum, as shown in Methods Section 3.4, the average displacement between consecutive frames should be 1 pixel or less. For recovering the orientation of the head, the affiliation of the roll angle to the different quadrants, as well as the sign determination of the pitch angle, rely on sufficient temporal sampling. For the former requirement, which is typically more demanding, the two extremums at 0° and 180° need to be detected for each full sperm head revolution. Thus, according to Nyquist's theorem, the minimal required framerate is defined as 4 frames per full revolution (360° roll) of the sperm head. Nevertheless, accuracy improves dramatically with much higher framerates. Specifically, the results obtained in this paper were taken at 2000 fps, and the sperm head rotated at 8 revolutions per second; thus we had approximately 250 frames per full head revolution. Another demand for tomographic reconstruction is an overall sufficient number of projections to properly fill the 3-D Fourier space of the object; yet this does not necessarily mandate a high framerate, and is thus is a weaker requirement. The results obtained in this paper used 1000 frames, taken over four full head revolutions.

For the task of 3-D localization of the flagellum, it is effectively treated as a superposition of point-like scatterers, which may reduce accuracy for 'singular' orientations of the flagellum relative to the camera, such as rod-like segments aligned along the optical axis. Sperm cells swimming towards the camera may also cause difficulties in recovering the head orientation, as in such cases the major radius may effectively turn into the minor radius of the projection and vice versa. Nevertheless, for sperm cells swimming in a perfusion chamber parallel to the camera, we have found that this scenario is rare.

In the analysis performed for the sperm head, we assume that the sperm cell head is a rigid body, which is a reasonable assumption due to a low confinement and flow rate (*20*), other than negligible deformation in the distal edge. We also assume that that the internal structure of the sperm head remained constant throughout the recording process, and only its orientation changed. In the results presented in this manuscript, the head RI reconstruction was based on four head revolutions, recorded over half a second, thus enabling the RI distribution to be updated twice every second during free swim.

Considering the fast acquisition rates, efficient digital implementation of the suggested algorithms has the potential to allow near real-time visualization. The head recovery process has to be performed once per sperm cell, using the first few hundreds of frames, in order to obtain the model constants and the 3-D RI reconstruction of the head. Afterwards, the orientation recovery in each frame is independent of all future frames, and can be done in real time. Regarding 3-D tomographic reconstruction of the head, we have previously shown that using a conventional computer graphic processing unit (GPU), the processing time for a full 3-D tomographic reconstruction neglecting diffraction, based on 73 256×256-pixel projections with known orientations, can be reduced to 0.007 seconds (*25*). In the current paper, we used 1000 128×128-pixel projections to reconstruct the sperm head 3-D RI, indicating that similarly fast processing rates would be possible using a GPU for this case. Full real-time implementations of the other parts of the algorithm, including flagellum 4-D segmentation and head focusing, are yet to be implemented.

To conclude, we presented a method enabling full 4-D reconstruction of freely swimming human sperm cells from an off-axis holographic video. In addition to assisting IVF procedures, by supplying quantitative 4-D morphological and dry mass evaluation of the sperm cells and their organelles, the method presented here might aid in new biological studies. Specifically, we believe that the elevated RI values detected near the midpiece area are attributed to the centriole region. This organelle, which is crucial for the development of the embryo as it is introduced into the embryo only by the sperm, could not be imaged in live cells until now. Future biological studies may verify our finding of increased RI in the centriole region, as well as the correlation between the identification of this region and successful fertilization.

**Materials and Methods**

1. Optical setup for interferometric phase microscopy

Figure S6 presents the optical setup used in this paper. Light from a Helium Neon laser source illuminates the sample in an inverted microscope, comprised of a 100× oil-immersion microscope objective, MO (Olympus UPLSAPO 100×O, numerical aperture: 1.4), and an achromatic tube lens, TL, of focal length 150 mm. The resulting sample beam then enters the off-axis external interferometric module (*26, 27*), where it is split into two beams of equal intensity by a 50:50 beam splitter, BS. The first beam is focused onto a laterally shifted retro-reflector mirror, RR1, by the first module achromatic lens, L1, of focal length 100 mm, causing a small shift in the illumination angle on the camera, producing off-axis interference. The second beam exiting the beam splitter is focused by lens L1 onto a 15 μm pinhole placed in the Fourier plane of the lens, thereby removing all the high spatial frequencies containing the sample information, thus creating a clean reference beam. The reference beam is then reflected back to BS by retro-reflector mirror RR2. The two beams then merge in the beam splitter and, after passing through another achromatic lens, L2 (focal length 150 mm) and an additional magnifying 4f system comprised of achromatic lenses L3 (focal length 30 mm) and L4 (focal length 75 mm), an off-axis image hologram is created on an ultra-fast digital camera (FASTCAM Mini AX200, Photron, square pixels of 20 μm each, 1024×1024 pixels, 2000 fps). The entire optical system possesses a total magnification of 328× and a resolution limit of 452 nm.

2. Biological preparation

A semen sample was collected in accordance with Tel Aviv University's institutional ethical committee, from a healthy 18 to 45 years old donor, after undergoing 24 hours of abstention. After ejaculation, the sample was allowed to liquefy for 30 min. Following this, sperm cells were isolated from the semen fluid using a PureCeption Bi-layer kit (Origio, Målov, Denmark) in accordance with manufacturer instructions. In short, 0.5 ml of semen were placed on top of a 40% and 80% silicon beads gradient and centrifuged for 25 min at 1750 RPM. After centrifugation at 1250 RPM for 5 min, the supernatant was discarded and the pellet containing the living sperm cells was washed with 10 ml of modified human tubal fluid (HTF) (Irvine Scientific, CA, USA). After additional centrifugation, the supernatant was removed and cells were resuspended in 2 ml of HTF supplemented with 7% Polyvinylpyrrolidone (PVP), 360,000 mw (PVP360, Sigma). 1 ml of the cell solution was then placed in a chamber (CoverWell, PC1L-0.5, 32-mm width × 19-mm length × 0.6 mm depth, 1.5-mm diameter ports) for imaging.

3. Algorithm

3.1. Wavefront propagation

An interferometric system is able to record the entire wavefront that propagated through a sample, rather than just its intensity, by interfering it with a reference wavefront, yielding a digital hologram or interferogram (*28*). The resultant hologram is given by the following expression:

$$|E_s(x,y)+E_r|^2 = |E_s(x,y)|^2 + |E_r|^2 + A_s(x,y)A_r \exp[j\varphi_s(x,y)-j\varphi_r] + A_s(x,y)A_r \exp[j\varphi_r - j\varphi_s(x,y)] \quad (1)$$

where *j* denotes the imaginary unit, and $E_s(x,y)=A_s(x,y)\exp[j\varphi_s(x,y)]$ and $E_r=A_r\exp[j\varphi_r]$ are the sample and reference complex wavefronts, respectively (the latter assumed to be of constant phase and amplitude). Thus, the wavefront that propagated through the sample is fully conserved, though its extraction remains difficult. In off-axis holography, one of the interfering beams is titled at a small angle relative to the other, creating a linear phase shift that allows separation of the field intensity from the two complex-conjugate wavefront cross-correlation terms in the spatial-frequency domain, thus allowing reconstruction of the complex sample wavefront from a single off-axis digital hologram. In this work, the complex wavefront was extracted from the raw hologram in the spatial-frequency domain using a Fourier-space filtering algorithm (*29*), which cropped one of the cross-correlation terms, and then applied an inverse Fourier transform to obtain the complex wavefront. For efficiency reasons, the images were left at their cropped dimensions, 4× smaller than the original hologram dimensions. Following this, the Rayleigh–Sommerfeld (RS) propagator was used to reconstruct the complex field at various distances from the recorded plane of focus (*30*), a propagation method suitable for weakly scattering objects (such as the sperm cell flagellum) (*31*). We took advantage of the transition to the Fourier space needed for isolating the wavefront, and used the Fourier formulation of the Rayleigh–Sommerfeld propagator, simply requiring a pixel-wise multiplication of the cropped cross-correlation term with the following transfer function (*30*):

$$H_{RS}(u,v,z=d) = \begin{cases} \exp\left[j\frac{2\pi n_m d}{\lambda}\sqrt{1-(\lambda u)^2-(\lambda v)^2}\right], & \text{for } \sqrt{u^2+v^2} < \frac{1}{\lambda}, \\ 0, & \text{otherwise} \end{cases} \qquad (2)$$

where $d$ is the distance to be propagated, $n_m$ is the RI of the surrounding medium, $j$ is the imaginary unit, and $\lambda$ is the illumination wavelength. Thus, the wavefront propagation may be regarded as a linear, dispersive spatial filter with a finite bandwidth. The refocused wavefront at $z=d$ can then be obtained by applying an inverse 2-D Fourier transform on the result. The spacing between the different reconstruction planes was chosen to be exactly the effective pixel size of the image (which was identical for horizontal and vertical coordinates), resulting in a pseudo-volume with isotropic sampling frequency. The boundaries of the propagation distance can be determined by the maximal depth allowed by the physical restriction used in the experiment. To prevent excessive memory use, a propagation distance of 16 μm is used in this paper, taking into consideration that the human flagellum is approximately 60 μm long, assuming that the sperm is not swimming towards the camera.

    3.2 Handling high noise levels

One of the greatest challenges of tracing the 3-D location of the flagellum per each frame is finding the pixels associated with the tail in the 2-D image, i.e. performing 2-D segmentation. This is a complex task due to the low phase values of the flagellum, which are similar to the noise level when using a coherent light source, especially for out-of-focus segments. We thus constructed a Targeted, Adaptable, Locally Conserving Cleaning (TALCC) function, which identifies the important parts of the image that need to be locally conserved. What makes this function highly targeted and adaptable is its multiple inputs, including not only the original phase image, but also the current binary segmentation map, row and column of interest, slope in region of interest, and noise level mode. The noise level mode, adapted automatically in our analysis, starts at a value of 0, indicating a neutral input, and can take higher positive values for dealing with increasingly higher noise levels that require robust cleaning, or lower negative values for dealing with thin, low-contrast segments that need delicate cleaning. The row and column of interest, together with the slope in region of interest, are used to define a 3×6 pixel environment around the row and column of interest that needs to be preserved, defined as one pixel to each side, 4 pixels forward and one backwards, rotated in the direction of the slope. The input phase map is first thresholded to obtain a binary image, with a threshold value respective to the noise level mode; then, both the pixel group defined by the 3×6 pixel environment and the pixel group marked as the object in the current (and previous frame – if available) segmentation map are marked as the object in the binary image. At this point, morphological opening and closing (*32*) are subsequently applied, in order to clean noise and close holes, respectively, with parameters according to the automatically detected noise level mode. The structuring elements used are lines with a slope either parallel or orthogonal to that of the region of interest, thus preserving segments with similar direction and erasing others. In the case of low-contrast segments, indicated by a negative noise level mode, the closing operation is applied prior to the cleaning operation, and the structuring elements used are disc-shaped.

### 3.3 Seed points

To aid the segmentation process in the low SNR conditions, we find a good ($x,y,z$) starting point in each frame, namely a "seed point", which is ideally the neck area connecting the sperm head and midpiece. Towards this end, all frames are first rotated according to the optical flow (*33*), such that the overall swimming direction of the cell is left to right. We start from the first frame, guessing the $z$ coordinate of the seed point to be at the middle of the range (i.e. the $z = 0$ plane). We then reconstruct the modulus of the phase profile at this initial depth, where the modulus operation is needed due to phase unwrapping problems that may occur in out-of-focus regions, and use it to calculate a binary mask of the head and midpiece areas. This is done by taking a low threshold (~0.25 radians), zeroing out most of the flagellum area, but not the slightly-elevated phase values of the midpiece area, and then applying morphological opening and closing. The orientation of the ellipse that has the same normalized second central moments as the binary mask is then calculated, indicating the estimated slope at the neck - $S_n$. The depth range for the search is determined by a fixed window parameter, which should be larger for the first frame, as the initial depth estimation is a guess, such that continuity is not expected. For each depth in the search area, a new binary mask of the sperm head and neck is calculated with a slightly higher threshold (~0.4 radians) as to exclude most of the flagellum, but not the midpiece. Then, the orientation of the ellipse that has the same normalized second central moments as the binary mask is calculated, and used to rotate the head to a horizontal position. In this horizontal position, morphological opening with a vertical-line structuring element is applied, effectively erasing the midpiece area with minimal damage to the rest of the cell, such that the leftmost column indicates the seed point. The length of the midpiece (used in Section 3.5) can then be evaluated by taking the difference between the original and cleaned rotated binary maps, and calculating the major axis length of the remaining object. After flagging this point, the mask can be rotated back to its original position, thereby determining the actual row and column of the seed point. Following this, a line centered in the $x$-$y$ location of the seed point and orthogonal to the neck, the direction of which was determined by $S_n$, is retrieved from the phase image cleaned by the TALCC function. This line is used to determine the ultimate focus location of the $x$-$y$ seed point by recovering numerous parameters. First, the line is fitted with a Gaussian, from which the width, expected value, amplitude and maximal value of the section are determined. Second, the number of maxima points in the original line values is determined. Finally, the sum of the absolute gradient of the line values in the amplitude image is calculated (*34*). Then, the smoothed width vector, amplitude vector, and sum of the absolute gradient vector, obtained for the entire depth range, are used to calculate a triple score vector, given by the element-wise multiplication of the sum of the absolute gradient and width vectors, divided by the amplitude vector. The depth with the lowest score – thus with the steepest, most narrow Gaussian, and with the most transparent amplitude image – corresponds to the ideal focus plane, yielding the depth of the seed point - $z_s$. To prevent poorly fitted Gaussians from corrupting the results, unreasonable parameters such as multiple maxima points, a negative amplitude, or an extremely high or low expected value all cause the respective score to be replaced with a high constant score, used as a flag. Third, the expected value retrieved from the Gaussian of the chosen depth is used to update the exact $(x_s, y_s)$ location of the seed point.

We repeat this process for all frames, where each time the initial $z$ coordinate of the seed point is the one found from the previous frame [$z_s(t-1)$], working under the assumption that the data is smooth enough due to the high framerate. Finally, we further utilize the expected smoothness between frames by applying a smoothing filter for each seed-coordinate vector: $\vec{x}_s$, $\vec{y}_s$, and $\vec{z}_s$.

3.4 Flagellum 4-D segmentation

In order to apply 3-D segmentation of the sperm cell flagellum, for each frame we iterate within a loop that begins with the seed point of that frame [$x_s(t)$, $y_s(t)$, $z_s(t)$], and runs until it reaches either the boundaries of the image or a stopping criterion. This means that this algorithm can only trace parts of the tail that are connected to each other, and not parts that exit the field of view and return in a secondary location. Within this loop, we run a recursive function (see detailed algorithm given in the Supplementary Information), which outputs the coordinates of the next voxel associated with the tail as well as a flag parameter, indicating whether we have reached the stopping criterion. This function takes as an input the coordinates of the last voxel that has been identified, the reconstructed complex field stack for that frame, two types of locations maps, an iteration counting parameter, the slope of the current segment, the noise level mode, and a validity flag variable initialized as a positive integer. The first type of location map stores non-zero values in $(x,y)$ locations identified as being associated with the flagellum, and the second type is based on the first type of the previous frame. Figure S7 shows two such location maps, where the map shown in A is the original binary location map restored from frame t, and the map shown in B is a weighted, dilated version of it, used for preventing deviation from the flagellum in frame $t+1$. For the first iteration of each frame, we use the coordinates of the seed point, the estimated slope at the neck ($S_n$) calculated in Section 3.3 and a neutral, zero noise level mode. For the first frame processed, we use an altered version of the function that does not require use of the second type of location maps.

The recursive function is composed of four main parts. In the first part, we evaluate the direction of the current segment, which is centered on the previous pixel ($x_o$, $y_o$) in the initial phase image. This initial phase image is reconstructed at depth $z_o$ (estimated for the previous pixel), and cleaned by pixel-wise multiplication with the weighted, dilated version of the location map of the previous frame. Next, this direction is used to define a directional environment to search the initial phase image for the coordinates of the next voxel associated with the tail, ($x_i$, $y_i$). In the second part, the $z_i$ location of ($x_i$, $y_i$) is recovered. This is achieved by first finding the possible range of depths based on continuity and smoothness considerations, both relative to the current and previous frame, and then examining a linear section of the respective depth phase image centered around ($x_i$, $y_i$), orthogonal to the direction of the current segment, for the calculated depth range. In this process, we calculate a score expressing the likelihood of every depth in this range to be the location of optimal focus, based on the distance from the center of the estimated depth range, the width, and the maximal phase value of the section. The validity of the score is then verified by a second set of parameters, including the estimated maximal phase value, width, standard deviation, and phase profile shape, in order to estimate whether the score is valid, more/less robust cleaning is required, or the end of the flagellum was reached. In the third part of the function, the validity score is evaluated, and used to decide whether to call the function again with a higher/lower noise level mode, finish the function normally, or finish the function with a flag indicating the end of the tail may have been detected. Finally, in the fourth part of the function, executed only once for the recursion base with the updated noise level mode value, we repeat the first part using the updated depth $z_N$, estimated at the second part. If the phase value of ($x_N$, $y_N$, $z_N$) is below a fixed threshold, we finish the function with a flag indicating the end of the tail may have been detected.

For each frame we iterate until we reach a stopping criterion, based on cumulative indications that the end of the flagellum may have been detected. If the flagellum detection has ended prior to exiting the field of view, we check the Euclidean distance between the current and previous end-point; if it is larger than a fixed threshold, we assume an error has occurred in this specific frame (as may occur due to high local noise), define this frame as invalid, and use the previous frame again instead, slightly increasing the distance tolerance for the next frame.

### 3.5 Flagellum integral RI evaluation

In order to calculate the integral RI value of the flagellum, we first classify all flagellum pixels into two groups: midpiece and non-midpiece. For this purpose, we evaluate the length of the midpiece at the detected seed point depth (see Section 3.3) in each frame. Based on the 4-D segmentation of the flagellum, we found that the average pitch angle of the midpiece is approximately 9°. Since the actual midpiece length is the length measured in the quantitative phase projection divided by the cosine of the pitch angle, the average error induced by not considering the pitch is only 1% of the midpiece length. Therefore, we have chosen to use the average midpiece length over all frames to classify the pixels identified as flagellum pixels according to Euclidian distance from the seed point. Note that a few dozen frames are enough to determine the length of the midpiece. To prevent bent tail positions from causing false association to the midpiece, once the first non-midpiece pixel is detected, all future pixels of that frame are declared non-midpiece as well. Next, the average of each group for the section width and maximal phase value parameters of the chosen depth, estimated in the second part of the recursive function described in Section 3.4, is calculated over all frames. Assuming the midpiece and non-midpiece portions of the flagellum can each be modeled as a cylinder, the thickness of the central pixel is equal to the width of the section; thus, the integral RI of each group can be calculated as follows:

$$n = n_m + \frac{\lambda A}{2\pi w \Delta x}, \quad (3)$$

where $n_m$ is the RI of the surrounding medium, $\lambda$ is the wavelength, $\Delta x$ is length of each pixel, calculated as the ratio between the sensor pixel size and the total magnification, $A$ is the average maximal phase value of the section for the group, and $w$ is the average section width in pixels for the group.

### 3.6 Head focusing

To retrieve the focus of the sperm head, we start from the first frame, guessing the optimal focus location of the head to be at the middle of the range (i.e. the $z = 0$ plane). We then reconstruct the modulus of the phase profile at this depth, and use it to calculate a binary mask of the head area. This is done by taking a low threshold (~0.25 radians), followed by morphological opening and closing. Then, the binary mask undergoes dilation, to include the area that will be occupied by the head in out of focus frames. The depth range for the search of optimal focus is determined by a fixed window parameter, which should be larger in value for the first frame, as the initial depth estimation is a guess, such that continuity is not expected. For each depth in the search area, we calculate: the sum of the absolute gradient of the amplitude image in the region defined by the binary mask (which should be minimal at the plane of focus for pure phase objects such as biological cells), the sum of the negative values of the phase image in the region defined by the binary mask (which is indicative of phase unwrapping issues characteristic of out of focus frames), and the Tamura coefficient [a sparsity metric (*35*)] of the gradient of the complex field in the region defined by the binary mask.

Following this, the smoothed parameter vectors obtained for the entire depth range are used to calculate a triple score vector, given by the element-wise multiplication of the first two, divided by the third. The depth with the lowest score corresponds to the ideal focus plane.

### 3.7 Head Segmentation

To achieve more precise segmentation, all quantitative phase images were resized to be 4× larger, returning them to the original hologram dimensions. For each frame, the segmentation of the sperm cell head from the original image consists of rough segmentation and fine segmentation. Initially, in the rough segmentation, the original image is segmented using a threshold of 0.6 radians on the phase values, followed by morphological closing and opening. This initial segmentation is used to estimate the centroid of the remaining object (which is mostly the head and possibly some of the midpiece), allowing the image to be cropped such that it is centered on the head. Following this, the orientation of the ellipse that has the same normalized second central moments as the binary mask is calculated, allowing the image to be rotated such that the head is placed horizontally, with the neck at the left and the acrosome at the right. Next, fine tuning of the segmentation is performed. At this step, we exploit the fact that the acrosome is in the right half of the image to apply more intensive, targeted cleaning to the left part, effectively and accurately erasing the midpiece area from the unsegmented phase image, leaving only the head. This is significant since for frames where the sperm cell is rotated on its side, the acrosome area becomes very thin at the midpiece area, and may be accidently erased if this cleaning is applied to the entire image. Thus, for the left third of the image, we apply a threshold of 0.9 radians on the phase values, followed by morphological opening with a structuring element that is a vertical line (unlike the usual disk-shaped structuring element), effectively erasing the midpiece area with minimal damage to the rest of the cell. For the right two thirds of the image, we only apply a much lower phase threshold of 0.35 radians. The resultant binary mask is then used to center and rotate the image more accurately, as well as to extract the minor and major axis radii of the ellipse that has the same normalized second central moments. Centering is performed for the bounding box enclosing the binary mask.

### 3.8 Head orientation recovery

In order to estimate the 3-D orientation of the sperm cell head for each frame, we model it to be an ellipsoid, according to:

$$(x/A)^2 + (y/B)^2 + (z/C)^2 = 1, \tag{4}$$

where it can be assumed from the standard dimensions of sperm cell heads that $B \gg A > C$ (see Fig. 1). This process finds the final orientation of the sperm head that yields the same projected ellipse as the rotated ellipsoid model, and not necessarily the stand-alone values of the pitch, roll and yaw angles, as 3-D rotation is not commutative (i.e., the order of rotation matters). The determination of the model constants $(A,B,C)$, as well as the affiliation of the pitch, roll and yaw angles to the different frames, relies completely on the minor and major axis radius vectors calculated for the binary masks of the respective projections of the head (as explained in Section 3.7). Thus, to prevent possible localized segmentation errors from corrupting the results, we apply a smoothing filter to each vector, relying on the smoothness that can be expected at high framerates. Examples of the major and minor radius vectors obtained for a sperm cell, before and after smoothing, are shown in Fig. S8, demonstrating the repetitive nature of sperm head rotation during free swim.

#### Finding the roll angle for each frame

We define the roll angle $\theta$ to be the rotation around the longitudinal axis of the sperm cell head ($y$ axis in Fig. 1). Due to the periodic nature of sperm head rolling during free swim, as demonstrated in Fig. S8A, constants *A* and *C* can be reliably estimated as the maximal and minimal values of the minor radius vector, respectively. Once constants *A* and *C* are known, and under the ellipsoid model assumption, the roll angle associated with each frame can be inferred within a 90° range from the minor radius of the projection. Practically, this can be done by calculating half the length of the projection of the ellipse shown in Fig 1 for a finite number of rotation angles in a 90° range, and then using interpolation to pinpoint the exact angle $\theta_{90}$ suitable for every measured minor radius length. To retrieve the angles over a 360° range, continuity considerations can be made in order to find the proper quadrant, as long as a high framerate is maintained. For example, we can label each segment between two extremums with a sequential integer $N_q$, and retrieve the angle within a 360° range by applying the following rule:

$$\theta_{360} = \begin{cases} \theta_{90}, & \text{if} \quad N_q \bmod 4 = 1 \\ 180 - \theta_{90}, & \text{if} \quad N_q \bmod 4 = 2 \\ 180 + \theta_{90}, & \text{if} \quad N_q \bmod 4 = 3 \\ 360 - \theta_{90}, & \text{if} \quad N_q \bmod 4 = 0 \end{cases}. \tag{5}$$

Note that relying on locating extremums defines the minimal required framerate to be four frames for a full revolution, yet accuracy improves dramatically with much higher framerates. For example, Fig. S8A shows approximately 250 frames per full revolution.

Finally, the direction of the rolling has to be addressed. The sperm cell can rotate either clockwise or counterclockwise, and can generally change its rotation direction during swim. Nevertheless, for the short recording duration needed for the reconstruction, we assume that the rotation direction is constant, a valid assumption for a smooth, periodic minor axis vector as the one shown in Fig. S8A. In order to automatically determine the directionality of the rotation for each sperm cell, we observe the 3-D vector indicating the location of the neck over time (the seed point found in Section 3.3). Specifically, assuming that the forward motion of the sperm is primarily in the *y* direction, we are interested in the *x* and *z* coordinate vectors. To determine rolling polarity, we calculate the gradient (numerical difference) of the *x* coordinate vector, $g_x$, and use it to calculate two parameters: $V_H$, which is the sum of all *z* coordinate values of frames for which $g_x > 0$, and $V_L$, which is the sum of all *z* coordinate values of frames for which $g_x < 0$. Then, if $V_H > V_L$, we determine the rotation direction to be clockwise, and vice versa. The logic behind this method is depicted in Fig. S8. If for a positive change in *x* (blue arrows) we obtain higher values of *z* than for a negative change in *x*, we are in the case depicted in Fig. S9A; otherwise, we are in the case depicted in Fig. S9B. Note that this calculation can be applied in an identical manner by switching the roles of *x* and *z* (and calculating $g_z$). If the rotation is counterclockwise, we simply need to correct the roll angles to be $\theta = -\theta_{360}$ Note that this calculation has to be performed only once for the first few dozen of frames, assuming that the cell does not change its rotation polarity.

**Finding the pitch angle for each frame**

We define the pitch angle $\chi$ to be the elevation of the head relative to the camera (*x-y*) plane (i.e. rotation around the *x* axis in Fig. 1). The pitch angle is related to the major axis radius, yielding a maximal value whenever the sperm cell head is aligned with the camera plane (i.e $\chi=0$). Thus, zero crossings of $\chi$, which are sign switching events, can be found from the global positive extremums of the major radius vector. Generally, for freely swimming sperm cells, the pitch angle displays a more complex rotation pattern than the roll angle. For example, the pattern exhibited in Fig. S8 shows a zero-crossing for $\chi$ after every full revolution in $\theta$. The correlation between the peaks in the major and minor radius vectors, where the global positive extremums of the major radius vector are a subset of the negative extremums of the minor radius vector, allows us to determine the zero crossings of $\chi$ in a more reliable manner that is less sensitive to noise. The missing constant of the model (*B*) can then be found from the average value of the global positive extremums in the major radius vector. Once all three constants are found, the pitch angle associated with each frame can be inferred from the major radius of the projection. Practically, this is done by rotating the 3-D ellipsoid according to the recovered roll angle $\theta$ for that frame, and then rotating it for a finite number of pitch angles within a 50° range, followed by calculating the major axis radius of each resultant projection. Then, interpolation is used to pinpoint the exact pitch angle value $\chi$ suitable for every measured major radius length. Note that a larger range than 50° is redundant as long as the primary forward motion of the sperm cell is in the *y* direction. If this is not the case, this method may be unsuitable, as the major radius may effectively turn into the minor radius of the projection for higher pitch angles.

Once the value of the pitch angle is found for all frames, we assign it with a proper sign, where each zero crossing of $\chi$ is expected to cause a sign switch, and we need to choose whether the even or odd segment numbers receive a negative sign. To determine this, we compare the *z* location vector of the neck, $z_{neck}$ (found in Section 3.3), to the *z* location vector of the head, $z_{head}$, as found in Section 3.6. Naively, we may expect $z_{head} > z_{neck}$ for $\chi>0$ and vice versa. Practically, however, this condition is not always fulfilled for every sample. Nevertheless, if we check the overall trend per segment between two zero crossings of $\chi$, it allows us to reliably estimate which segments are positive and which are negative.

**Finding the yaw angle for each frame**

We define the yaw angle $\varphi$ to be the rotation in the camera (*x*-*y*) plane (i.e. rotation around the *z* axis in Fig. 1). The recorded yaw angle $\varphi_r$ has already been found from the orientation of the ellipse that has the same normalized second central moments as the binary mask in Section 3.7, and should not be an input to tomographic reconstruction, as we have already rotated all sperm cell heads in -$\varphi_r$ to achieve a horizontal position. However, for non-spherical objects, the combination of non-zero roll and pitch inevitably creates a small yaw angle $\varphi_0$, which should be calculated and compensated when preparing the phase image to be located in the native position for the tomographic reconstruction. This small angle can be found for each frame by rotating the 3-D ellipsoid model according to the previously calculated roll and pitch angles, and then calculating the orientation of the ellipse that has the same normalized second central moments as the binary mask of the resultant projection. Note that for recovering the 3-D orientation of the sperm cell head for movement reconstruction, we impose a yaw angle of $\varphi_r - \varphi_0$, mimicking the measured yaw angle.

3.9 Tomographic reconstruction

Optical diffraction tomography (ODT) is a well-established technique, which enables reconstruction of the 3-D RI distribution of an object from a set of digital holograms taken from multiple viewing angles of the object. Specifically, the Rytov approximation for the diffraction algorithm is valid as long as the phase gradient in the sample is small (*22*, *36*, *37*). When applying the ODT algorithm, the Rytov field for each properly rotated and centered hologram is calculated as follows:

$$u_{Rytov}(x, y) = \log(|E_s(x, y)|/|E_0(x, y)|) + j\varphi(x, y), \qquad (6)$$

where $|E_s(x,y)|$ is the amplitude extracted from the sample hologram, $|E_0(x,y)|$ is the amplitude extracted from a sample-free acquisition taken as a reference, $\varphi(x,y)$ is the difference between the phase profile extracted from the sample hologram and the phase profile extracted from a sample-free acquisition, and *j* is the imaginary unit. The 2-D Fourier transform of the Rytov field for each hologram taken from viewpoint ($\theta,\chi,\varphi$) is then mapped into a 2-D hemispheric surface (Ewald sphere) rotated in the same direction in the 3-D spatial Fourier space (*22*). In the case discussed here, where there is no constant rotation axis that can be defined as one of the Cartesian axes, the available sampling points in the 3-D frequency domain are naturally unevenly distributed in the 3-D Cartesian grid, causing significant artifacts. This can be partially resolved by an additional interpolating step for estimating the unknown spectral values on a predefined 3-D Cartesian grid (*38*). For this purpose, we used the Inverse Distance Weighting (IDW) interpolation method (*39*), which is based on the assumption that the frequency sample that is to be interpolated should be more strongly influenced by closely located neighboring samples than by remotely located neighboring samples, such that the value at the desired location is a weighted linear average of the neighboring values, where the associated weight decreases with distance. This can be mathematically formulated as follows:

$$F_{cartesian} = \sum_{i=1}^{N} w_i \cdot f_i, \quad w_i = \frac{h_i^{-P}}{\sum_{k=1}^{N} h_k^{-P}}, \tag{7}$$

where $f_i$ represents the spectral value of the $i$-th scattered frequency sample and $w_i$ is its associated weight, calculated based on the relative inverse distance between the weighted sample and the interpolation site. $N$ and $P$ are two significant parameters of this method, where $N$ is the number of neighbors considered, and $P$ is the weighting factor, where greater values of $P$ assign greater influence to the values closest to the interpolated point. We chose $N \geq 30$, implemented by looking for neighbors in a growing 3-D cube until at least 30 neighbors are encountered, and $P=4$, according to a rule of thumb stating that for a $D$-dimensional problem, $P \leq D$ causes the interpolated values to be dominated by points far from the interpolation site.

Finally, the 3-D RI distribution of the original object can be obtained by performing an inverse 3-D Fourier transform, and applying the following formula:

$$n(x, y, z) = \sqrt{n_m^2 - \frac{\lambda^2}{4\pi^2} \cdot f(x, y, z)}, \tag{8}$$

where $n_m$ is the RI of the medium, $\lambda$ is the illumination wavelength, and $f(x,y,z)$ is the result of the inverse 3-D Fourier transform. Note that the quality of tomographic reconstruction improves with the available number of viewpoints, where we used 1000 projections.

**Acknowledgments**

**Funding:** This work was supported by the H2020 European Research Council (ERC) grant (678316), the Clore Israel Foundation, and the Tel Aviv University Center for Light-Matter Interaction.

**Author contributions:** N.T.S., G.D.Y., and S.K.M. conceived the idea. G.D.Y. and N.T.S. developed the algorithms. G.D.Y. implemented the algorithms, performed the numerical simulations and processed the experimental data. S.K.M. performed the optical recording of the sperm cells. I.B. performed the biological preparation of the sperm cells used in this paper, wrote the biological protocol, and supplied biological insights. G.D.Y., S.M.K. and N.T.S. wrote the paper. N.T.S. supervised the project.

**Competing interests:** We declare that there are no competing interests.

**Data and materials availability:** The raw and analyzed datasets generated during this study are available for research purposes from the corresponding author at nshaked@tau.ac.il on reasonable request.

**Figures and Tables**

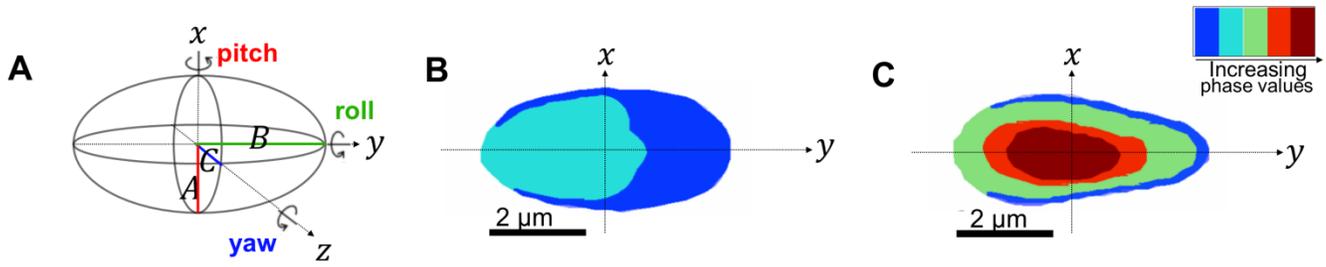

**Fig. 1. Schematic illustrations for sperm cell head orientation recovery algorithm, where the camera plane is parallel to the *x-y* plane.** (**A**) 3-D ellipsoid model, including definition of the pitch, roll, and yaw angles, as well as the three model constants, A, B, C. (**B**) The projection of a sperm cell head lying flat. (**C**) The projection of a sperm cell head lying on its side.

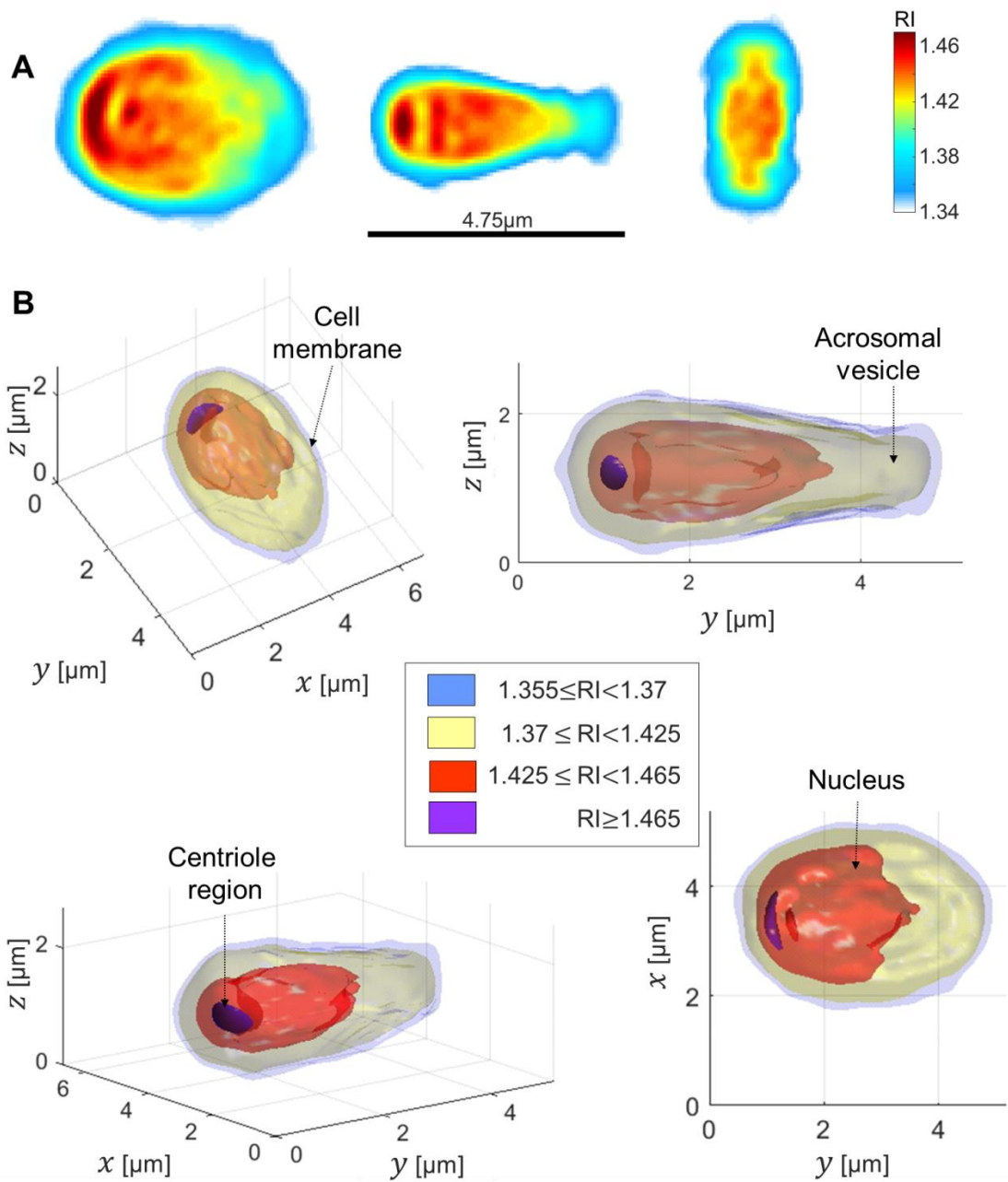

**Fig. 2. A representative sperm cell head 3-D RI reconstruction from 1000 frames during free swim.** (**A**) Middle slices along the three main axes, colorbar denotes RI values. (**B**) Thresholded 3-D rendering from various perspectives. Two other cells are shown in Fig. S5

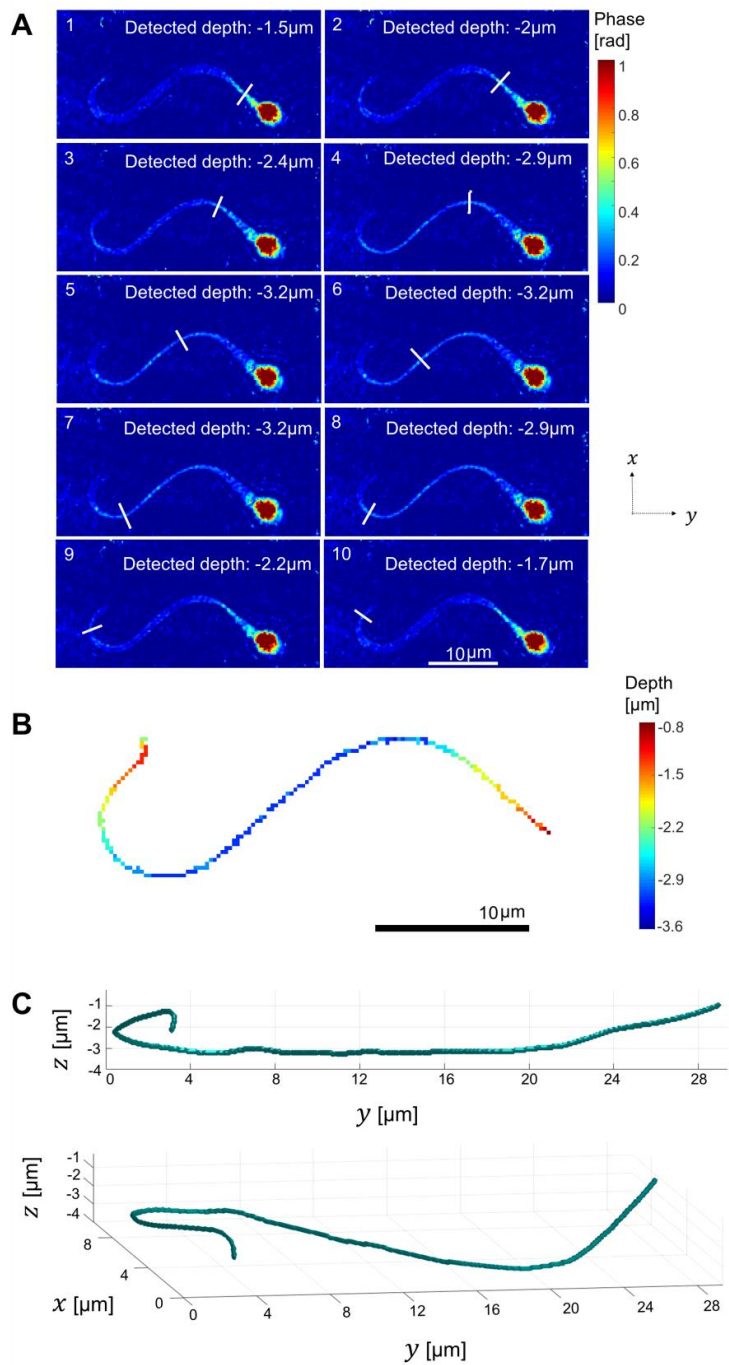

**Fig 3. 3-D dynamic reconstruction of a human sperm flagellum during free swim.** (**A**) Digital focusing in action. The white segment moves along the flagellum from the neck to the distal end, in the order indicated by the numbers in the top-left corners, finding the ultimate focus at each location. Colorbar indicates phase values in radians. All the images shown are retrieved from a single digital off-axis hologram, acquired in a single camera shot. See dynamic representation in Movie S1. (**B**) 2-D representation of the 3-D segmentation, where the colormap indicates the recovered depth (z coordinate). (**C**) 3-D segmentation result from two viewpoints. See dynamic representation in Movie S2.

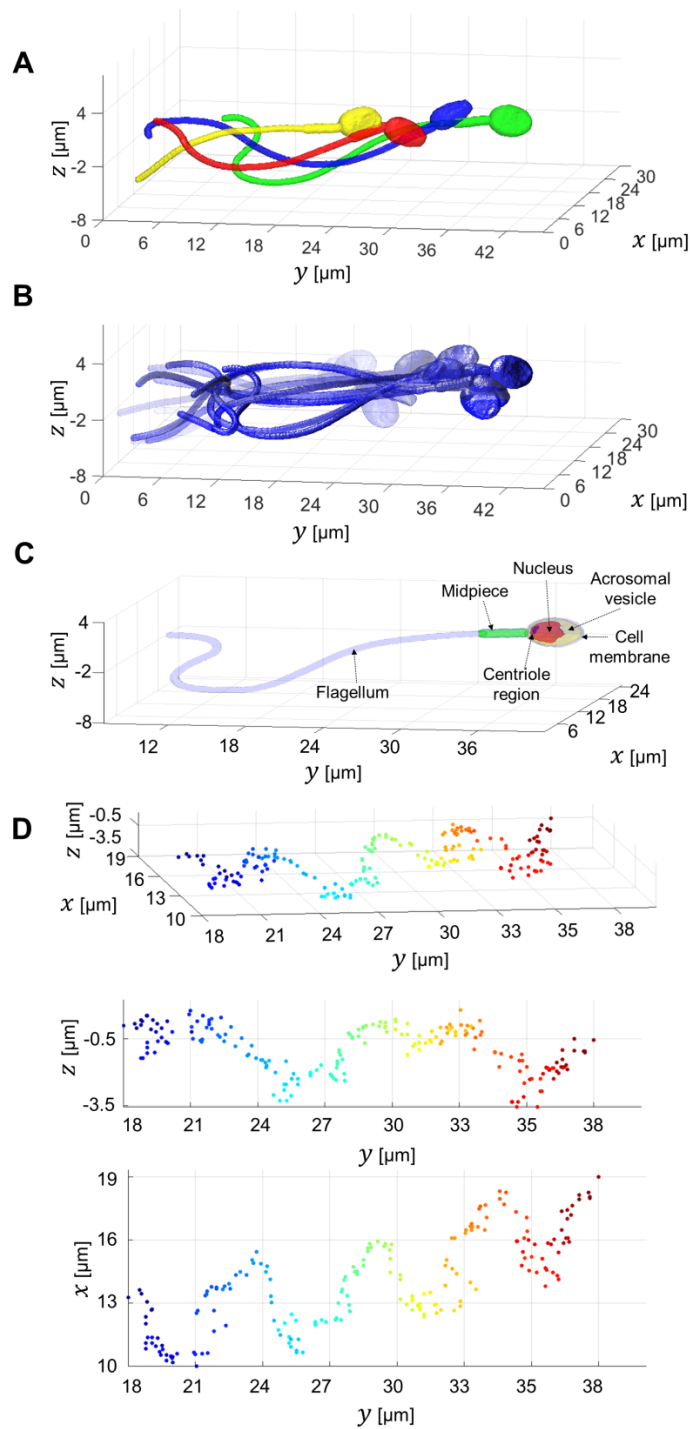

**Fig 4. Full 4-D reconstruction of a sperm cell, obtained from 1000 frames recorded during free swim.** (**A**) Overlay of four frames from the 3-D motion in Movie S3, each assigned with a different color. (**B**) Overlay of 15 frames from the 3-D motion in Movie S3, each assigned with a different opacity, where earlier times are more transparent. (**C**) A single frame from the 3-D motion in Movie S3, revealing the internal structure of the sperm cell. Light purple indicates the cell membrane ($1.355 \leq RI < 1.37$), green indicates the midpiece ($RI = 1.383$), yellow indicates the acrosomal vesicle ($1.37 \leq RI < 1.425$), red indicated the nucleus ($1.425 \leq RI < 1.465$), and dark purple indicates $RI \geq 1.465$ (centriole region). (**D**) Different views of the 3-D head centroid trajectory, changing from blue to red as time progresses.

**Supplementary Materials**

*Figures*

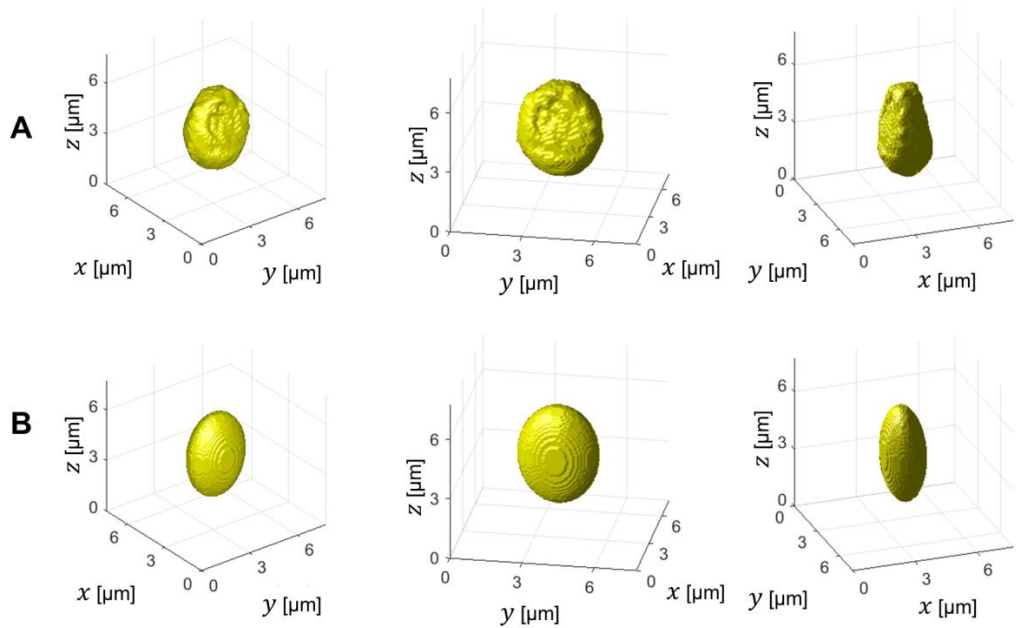

**Fig. S1. 3-D surface plot of an actual sperm cell (A) vs. its ellipsoid model (B), from various viewpoints.** The actual shape of the sperm cell is based on the results of the tomographic reconstruction obtained for experimental data (Fig. 2 in the main text), and coincides with the predicted shape of the head. Unlike the inner organelles of the sperm cell head, the 3-D shape of the sperm cell head surface is well known. The ellipsoid model in B was calculated based on the holograms of the same cell, as explained in Methods Section 3.8.

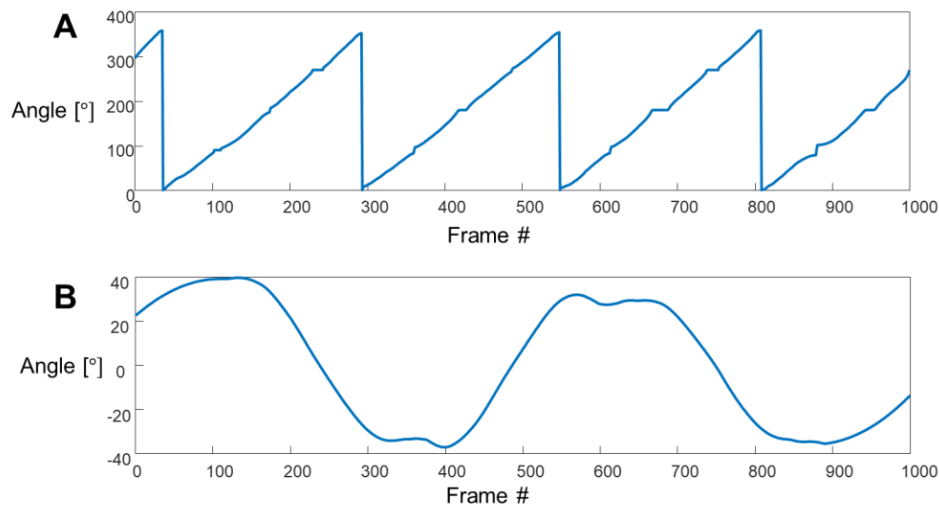

**Fig. S2. Experimentally recovered orientation of a human sperm cell during free swim.** (**A**) Roll angle. (**B**) Pitch angle.

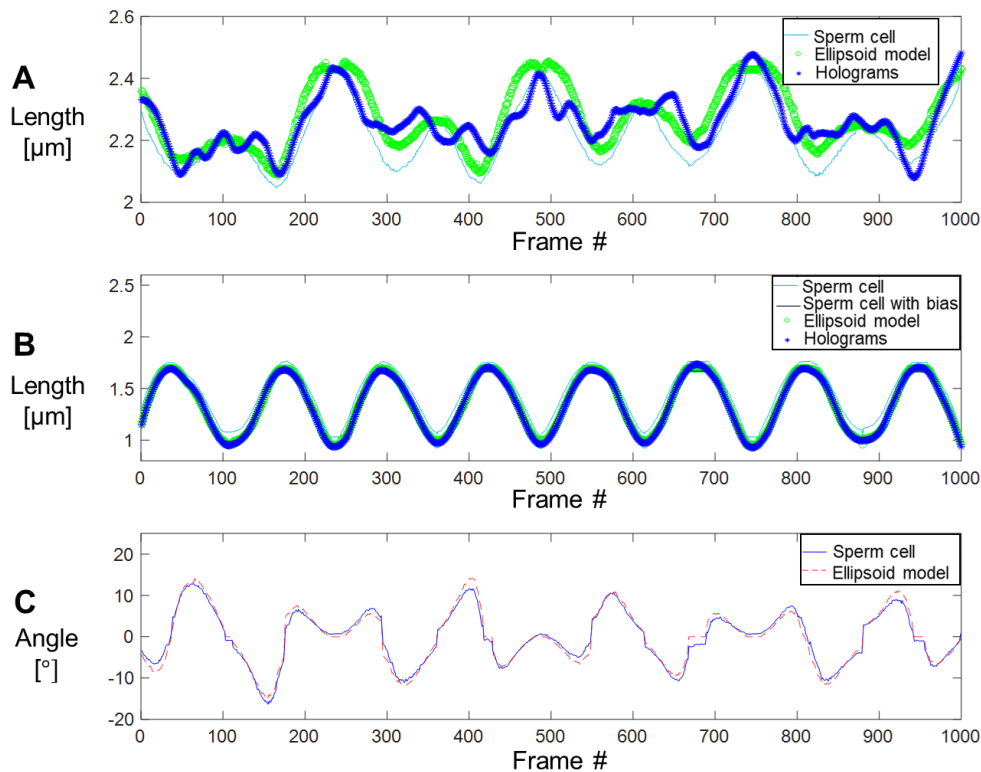

**Fig. S3. Comparison of various parameters calculated for the actual sperm cell head and the ellipsoid model, demonstrating negligible errors in calculating head orientation based on this model.** (**A**) Major radius. The root mean squared error (RMSE) is 0.069 μm between the results obtained from the sperm cell and ellipsoid model, and 0.07 μm between the result obtained from the sperm cell and the holograms. (**B**) Minor radius. The RMSE is 0.1 μm both between the results obtained from the sperm cell and the ellipsoid model, and between the results obtained from the sperm cell and the holograms. Here, we also show the result obtained for the sperm cell when adding a constant bias of 1.5 pixels to correct segmentation errors. After removing the constant bias, the RMSE is reduced to 0.03 μm between the result obtained from the sperm cell and the ellipsoid model, and 0.034 μm between the result obtained from the sperm cell and the holograms. (**C**) Yaw angle. The RMSE between the result obtained from the sperm cell and the ellipsoid model is 1.16°. The native yaw angle shown here cannot be reconstructed from the hologram itself, thus this third curve is not shown here.

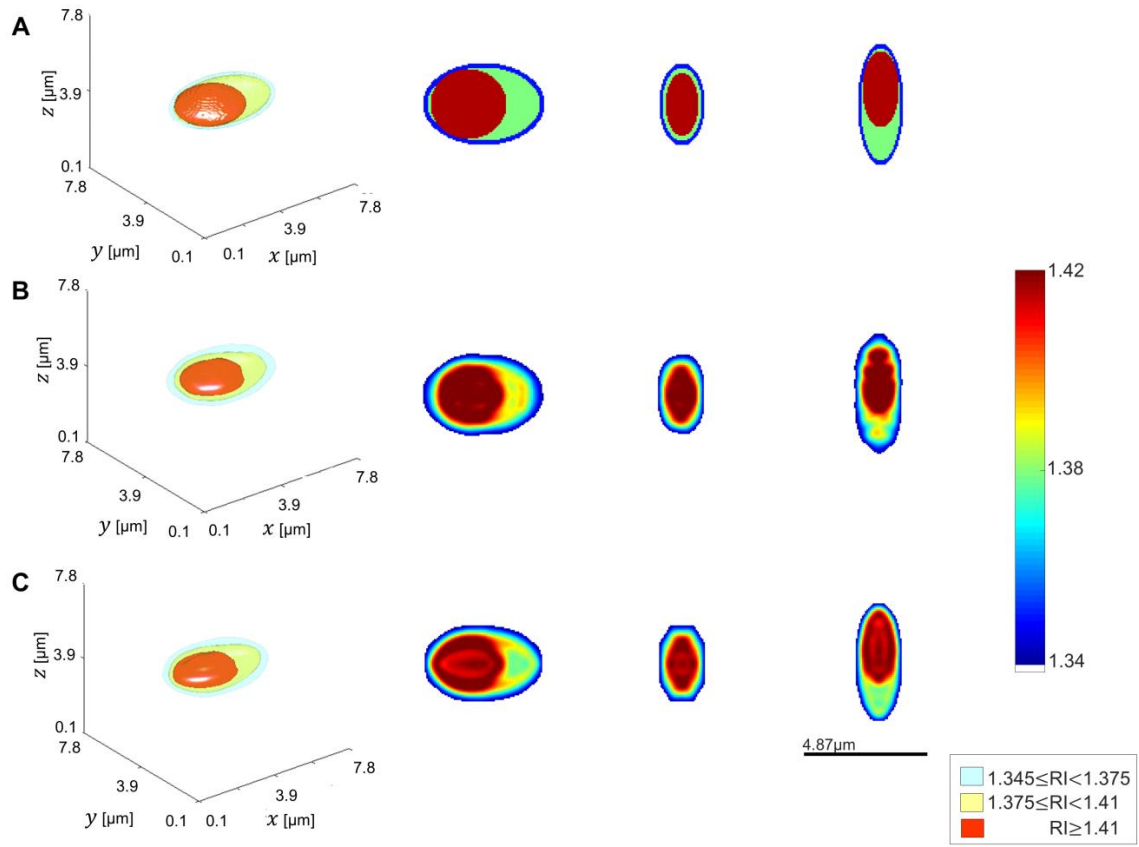

**Fig. S4. Simulation results verifying that the exact set of orientations obtained from the experimental data of the freely swimming sperm can obtain high-quality tomographic reconstruction.** (**A**) Ground-truth model of the RI distribution of a sperm head. (**B**) Reconstructed RI distribution obtained by using exactly the same orientations that have been available in the experimental case. (**C**) Reconstruction of RI distribution model using a single-axis 360° rotation at 5° angular increments, showing almost identical reconstruction quality as in B. In the leftmost column, the 3-D rendering is shown, with the corresponding legend provided in the bottom right corner. In the second column a mid x-y section is shown. In the third column a mid y-z section is shown. In the fourth column a mid x-z section is shown. The colorbar on the right displays RI values for the slice images.

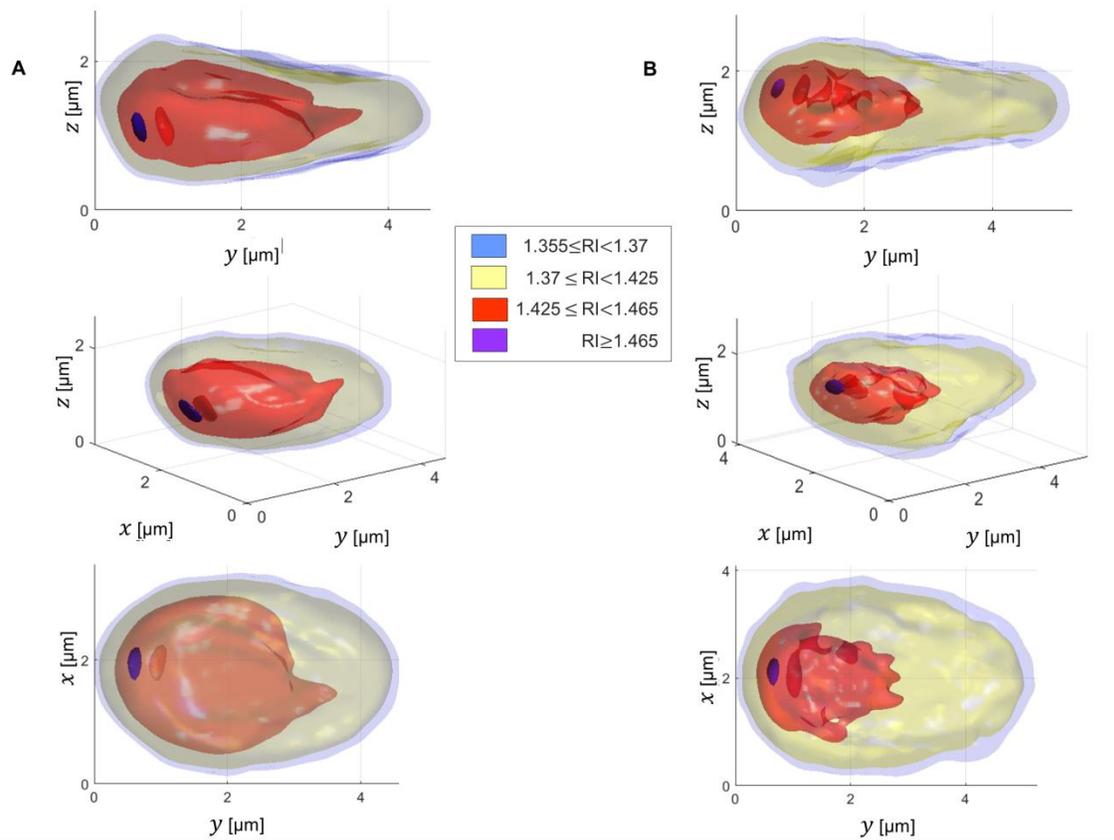

**Fig. S5. 3-D RI reconstructions of two additional sperm heads during free swim.** (**A**) Reconstruction based on 1000 frames, recorded at 2000 fps. (**B**) Reconstruction based on 500 frames, recorded at 1000 fps.

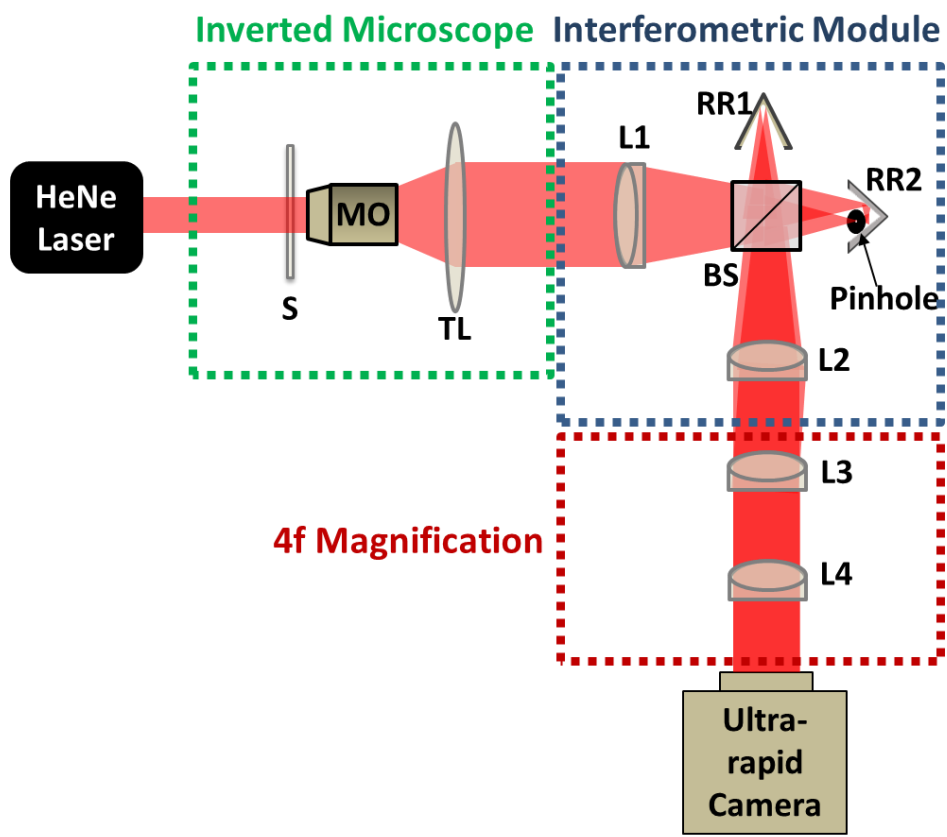

**Fig. S6. Diagram of optical system used to acquire the quantitative phase maps of the freely swimming sperm.** S, sample; MO, microscope objective; TL, tube lens; L1 - L4, achromatic lenses; BS, beam splitter; RR1, RR2, retro-reflector mirrors.

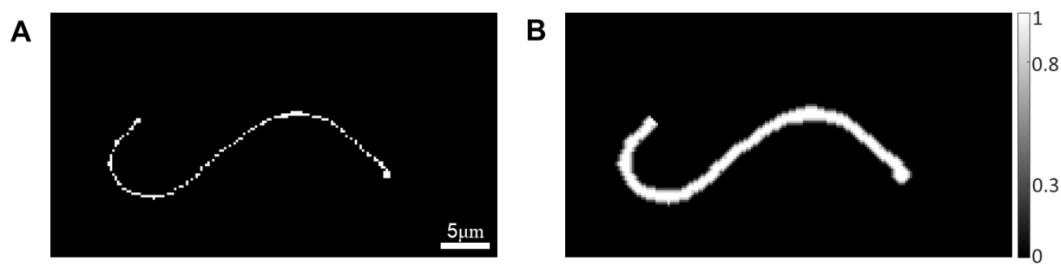

**Fig. S7. An example of the binary and weighting maps used for the recursive function (findFOC), for the frame shown in Fig. 3 in the main text.** (**A**) OLDMAP (**B**) OLDMAP2.

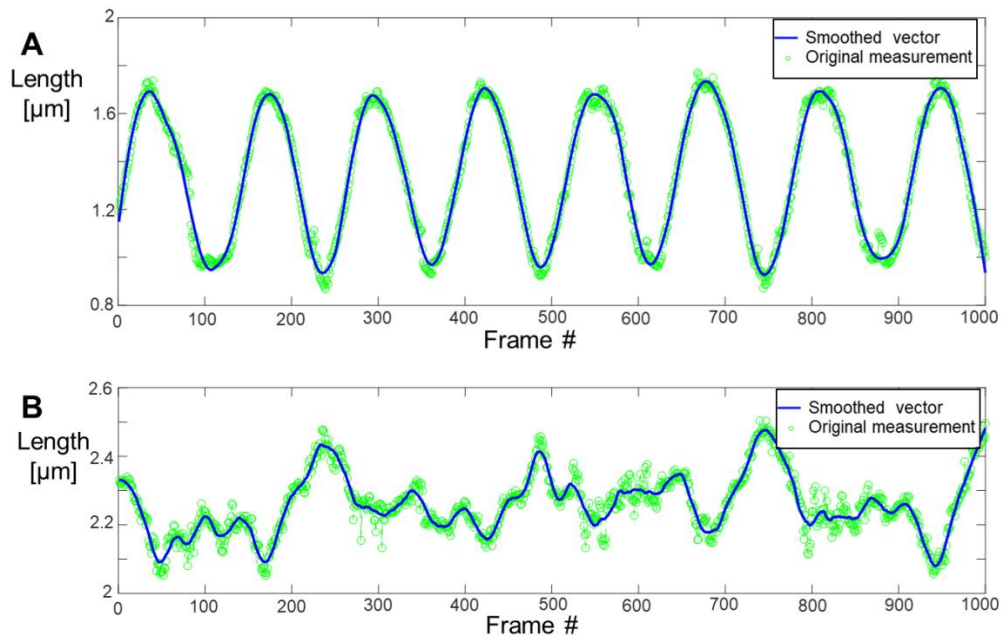

**Fig. S8. Major and minor radius measured for the experimental data, before and after smoothing.** (**A**) Minor radius. (**B**) Major radius.

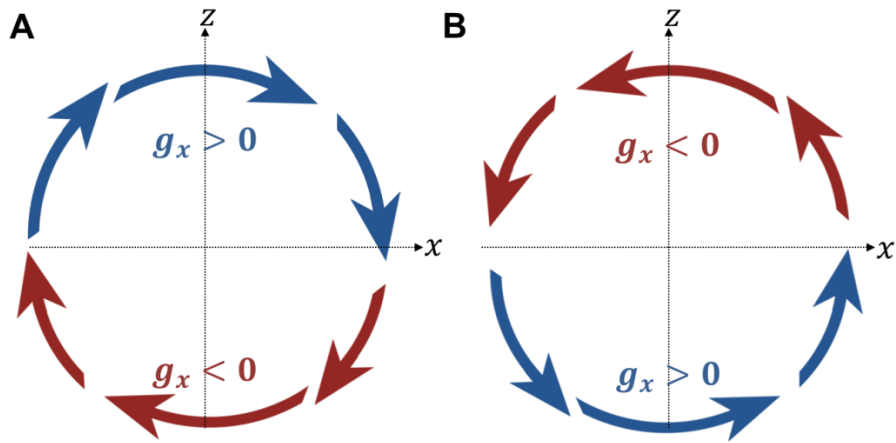

**Fig. S9. Method of finding the sperm head roll direction.** Red arrows indicate $g_x < 0$, and blue arrows indicate $g_x > 0$. (**A**) Clockwise: higher values of z are obtained for the blue arrows ($g_x > 0$). (**B**) Counterclockwise: higher values of z are obtained for the red arrows ($g_x < 0$).

*Materials and Methods*

**Detailed algorithm (psudo-code) for processing the flagellum**

The recursive function, named findFOC, takes as input the coordinates of the last voxel that has been identified, the reconstructed complex field stack for that frame, six location maps (hit, hitc, hitz, OLDMAP, OLDMAPz, and OLDMAP2), an iteration counting parameter, the slope of the current segment, the noise level mode, and a validity flag variable initialized as a positive integer. The hit, hitc and hitz location maps store non-zero values in ($x$,$y$) locations identified as being associated with the flagellum for the current frame, where the first stores consecutive integers that increase as the loop progresses (iteration count), the second stores "1"s, and the third stores the recovered depth for that location. OLDMAP is simply the hitc map for the previous frame, OLDMAPz is the hitz map for the previous frame, and OLDMAP2 is a dilated version of OLDMAP. OLDMAP2 is generated by applying multiple morphological dilations to OLDMAP with a disk-shaped structuring element 1 pixel in size, where pixels added by later dilations get a lower weight, yielding a weighting object that relies on smoothness (see Fig. S7) to indicate the likelihood of finding sperm pixels. In order to keep the endpoint in OLDMAP2 intact, we then zero out a 7×5 pixel environment around the row and column detected as associated with the end of the flagellum, defined as 3 pixels to each side, and 5 pixels forward, rotated at the direction of the slope of the last detected segment. For the first iteration of each frame, we use the coordinates of the seed point, the estimated slope at the neck ($S_n$) calculated in Section 3.3, and a neutral (i.e., 0) noise level mode. For the first frame processed, we use an altered version of the function (findFOC0) that does not require OLDMAP and OLDMAP2.

The findFOC function is composed of four parts:

1. **Initial estimation of current segment direction & ($x_i$,$y_i$) coordinates of the next voxel associated with the flagellum.** In this part, we calculate the phase profile for the depth estimated from the previous voxel ($z_o$), and clean it by multiplying it by the weighting map OLDMAP2 (for findFOC0: using the TALCC function, with the parameters calculated from the previous iteration). Then, we estimate the slope at the previous location ($x_o$,$y_o$); if the iteration counting parameter is lower than a fixed threshold, indicating that we are near the head, the slope is calculated based on thresholding the phase image to leave the head, and taking the orientation of the ellipse that has the same normalized second central moments as its binary mask. Otherwise, we crop the OLDMAP (for findFOC0: hitc) binary mask using a fixed window size centered on ($x_o$,$y_o$), and estimate the orientation of the remaining object. This procedure only gives us the non-unique slope within a 180° range: -90°→90°; to find the slope within a 360° range, which is crucial for moving forward with the flagellum rather than backwards, we calculate the gradient of the hit location map (rotated according to the non-unique slope), the sign of which indicates the directionality of the search. To add to the reliability of the procedure, we assume continuity in the slope relative to the previous estimation, such that calculated slopes that are different by more than a fixed threshold from the previously calculated slope are discarded and replaced with the former result. This calculated slope, namely $S_N$, is used to define a 3×2 pixel environment centered on ($x_o$,$y_o$), defined as 1 pixel to each side, and 1 pixel forward, rotated in the direction of the slope. Finally, we search the clean phase profile in the calculated environment for the maximal value, excluding pixels already flagged as visited by the hitc location map (which are flagged as "0"s), where the location of the maximal value yields the initial estimation of the ($x_i$,$y_i$) coordinates of the next voxel associated with the flagellum. If all values in the environment are "0", indicating that all locations in the environment have been visited, we define a validity flag of 0, indicating that we have reached the end of the flagellum for this frame (stopping criterion), and return to the calling function immediately.

2. **Estimation of the $z_N$ location of the ($x_i,y_i$) coordinate.** In this part, we first find the possible range of depths based on continuity and smoothness considerations, both relative to the current and previous frame (meaning in 4-D); towards this end, we first extract the depth range from a 5×5 pixel environment in the OLDMAPz map, defined as 2 pixels to each side, and 4 pixels forward, rotated in the direction of the slope, starting at ($x_i,y_i$), and calculate its median, $z_m$. We then use this information as follows:

- If $z_m < z_o$, we take [$z_o$ -2, $z_o$].
- If $z_m > z_o$, we take [$z_o$, $z_o$ +2].
- If $z_m = z_o$ (or there is insufficient information in the area defined by the 5X5 pixel environment in the OLDMAPz map), we simply take [$z_o$ -2, $z_o$ +2].

Next, we examine a linear section of the TALCC-cleaned phase image centered on ($x_i,y_i$) which is orthogonal to the direction of the current segment (based on $S_N$), for the calculated depth range. Since the midpiece area is characterized by a thicker segment, such that it can endure more rigorous cleaning on the one hand, and is less sensitive to our noise detection criteria on the other hand, by default we use a higher noise level mode when cleaning it with the TALCC function (see Section 3.5). For every depth in the calculated range, we calculate the standard deviation and a score expressing the likelihood of this depth being the location of optimal focus, based on the following parameters:

1- Weighting according to distance from the center of the estimated depth range. Assuming continuity and smoothness, it is most likely that the depth of the following pixel will be identical to the current one, slightly less likely for it to differ by ±1, and increasingly less likely as the difference increases.
2- The width of the section (number of pixels with value above threshold).
3- The amplitude of the section (value of highest peak).

To minimize the effect of local noise, we smoothed the width and amplitude vectors with a window size equal to the vector length. Next, the distance-weighting, width, and amplitude-weighting vectors obtained for the entire depth range are used to calculate a triple-score-vector, given by the element-wise multiplication of the distance-weighting and width vector, divided (element-wise) by the amplitude vector. At this point, we need to check the validity of the score, which is linked to the noise level of the image. If the image is noisy even after the TALCC function, the width parameter would be unreliable, as it may include background pixels. On the other hand, if the cleaning procedure was too aggressive, we may have excluded pixels that should have been counted. Thus, the score is flagged by replacing it with a large constant number in the following cases:

- The amplitude value is 0.
- The segment has no maximum points (shape is wrong).
- The width of the segment is higher than a threshold (set to a different value in the thicker midpiece area than in the narrow flagellum area).

- The width of the segment is two pixels or less, and the width of the respective segments in adjacent depths differs by over two pixels (suggesting a mistake in segmentation).
- An additional condition only relevant for non-midpiece area: the segment has minimum points (i.e., the shape is wrong).

Afterwards, the closest neighbors of the depths that received the large constant number have their scores doubled as a penalty, with the assumption being that it is less likely for a depth adjacent of these invalid depths to be the ideal focus plane.

Then, the depth with the lowest score – thus with the steepest, most narrow segment (that did not fall into any of the categories above) – corresponds to the ideal focus plane, yielding the depth of the current location, $z_N$. If the value of the lowest score is the high constant number used as a flag, we infer that the noise level is still too high and return a "0" noise flag to the calling function. We also return a "0" noise flag (indicating high noise levels) to the calling function if we are out of the midpiece area with a noise level mode lower than three (meaning more rigorous cleaning was not applied) and there is either a jump of more than two pixels in the width vector (indicating unrelated pixels attributed to the count), or more than one depth with an invalid shape.

Finally, we calculate two additional global parameters to return to the calling function, indicating either excessive cleaning or that the end of the flagellum was reached:

1- The mAMP parameter, storing the maximal value of the amplitude vector.
2- The mSTD parameter, storing the maximal value of the standard deviation vector (calculated for the raw, uncleaned phase image).

3. **Validity check of the result and recursion**

In this part, we look at the output of the previous part and decide whether the result is valid, more/less cleaning is needed, or we have reached the end of the flagellum.

- If the current noise level mode is non-negative (indicating neutral or high noise) but smaller than 5 (to prevent an infinite loop) and the noise flag from the previous step is 0: we recall the findFOC function with a noise level mode increased by 1, with all the initial parameters, including $(x_o, y_o, z_0)$.
- If the current noise level is non-positive (indicating neutral noise or delicate segments) but larger than −3 (to prevent an infinite loop), and mAMP or mSTD are lower than a fixed threshold: we recall the findFOC function with a noise level mode decreased by 1, with all the initial parameters, including $(x_o, y_o, z_0)$.
- If none of the above conditions were met and mAMP or mSTD are lower than a fixed threshold, we decrease the value of the validity flag by 1.

4. **Final estimation of the current segment direction and the $(x_N, y_N)$ coordinates of the next voxel associated with the flagellum.** In this part, executed only once for the recursion base, with the updated noise level mode value, we repeat step 1 using the updated depth $z_N$, estimated at Step 2 above. If the phase value of $(x_N, y_N, z_N)$ is below a fixed threshold, we decrease the value of the validity flag by 1 (unless it has already been decreased by 1 in this iteration).

For each frame, we iterate until we reach a value of 0 for the validity flag, indicating that we have reached the end of the flagellum for this frame (stopping criterion). The tolerance for local errors vs. the ability to accurately pinpoint the end-location of the flagellum is a result of the positive integer chosen for the initialization of the validity flag variable. If the flagellum detection has ended abruptly by reaching a validity flag of 0, we check the Euclidean distance between the current and previous end-point; if it is larger than a fixed threshold, we assume an error has occurred in this specific frame (as may occur due to high local noise), define this frame as invalid, and use the previous frame again instead, slightly increasing the distance tolerance for the next frame.

**Movie S1**. **Focusing algorithm in action**. The red vertical segment moves along the flagellum from the neck to the distal end, finding the ultimate focus plane for each location it is in.

**Movie S2. 2-D representation of the 3-D segmentation map of the flagellum.** The colormap indicates the recovered depth (relative to the original focus plane).

**Movie S3. Full 3-D motion reconstruction of a human sperm cell swimming freely.**

**These movies are available here:** http://www.eng.tau.ac.il/~omni/index2.php?p=6704